\documentclass[aps,prl,reprint,superscriptaddress,showpacs,longbibliography,footnoteinbib]{revtex4-1}
\usepackage[utf8]{inputenc}
\usepackage{xcolor}
\usepackage{pdfcolmk}
\usepackage{float}
\usepackage{bm}
\usepackage{amsmath}
\usepackage{amssymb}
\usepackage{graphicx}
\PassOptionsToPackage{normalem}{ulem}
\usepackage{ulem}
\usepackage[unicode=true,
 bookmarks=false,
 breaklinks=false,pdfborder={0 0 1},backref=false,colorlinks=true]
 {hyperref}
\hypersetup{pdftitle={Title},
 plainpages=false,pdfpagelabels,linkcolor=blue,urlcolor=blue,citecolor=blue,pdfdisplaydoctitle=true,pdfduplex=DuplexFlipLongEdge}

\makeatletter

\providecolor{lyxadded}{rgb}{0,0,1}
\providecolor{lyxdeleted}{rgb}{1,0,0}

\DeclareRobustCommand{\lyxsout}[1]{\ifx\\#1\else\sout{#1}\fi}

\usepackage{units}\usepackage{wasysym}

\definecolor{orange}{rgb}{0.80, 0.50, 0.0}

\usepackage{bm}
\usepackage{braket}

\makeatother

\begin{document}
\noindent\begin{minipage}[t]{1\columnwidth}%
\global\long\def\ket#1{\left| #1\right\rangle }%

\global\long\def\bra#1{\left\langle #1 \right|}%

\global\long\def\kket#1{\left\Vert #1\right\rangle }%

\global\long\def\bbra#1{\left\langle #1\right\Vert }%

\global\long\def\braket#1#2{\left\langle #1\right. \left| #2 \right\rangle }%

\global\long\def\bbrakket#1#2{\left\langle #1\right. \left\Vert #2\right\rangle }%

\global\long\def\av#1{\left\langle #1 \right\rangle }%

\global\long\def\tr{\text{tr}}%

\global\long\def\Tr{\text{Tr}}%

\global\long\def\pd{\partial}%

\global\long\def\im{\text{Im}}%

\global\long\def\re{\text{Re}}%

\global\long\def\sgn{\text{sgn}}%

\global\long\def\Det{\text{Det}}%

\global\long\def\abs#1{\left|#1\right|}%

\global\long\def\up{\uparrow}%

\global\long\def\down{\downarrow}%

\global\long\def\vc#1{\mathbf{#1}}%

\global\long\def\bs#1{\boldsymbol{#1}}%

\global\long\def\t#1{\text{#1}}%
\end{minipage}
\title{Quasiperiodicity hinders ergodic Floquet eigenstates}
\author{Miguel Gonçalves}
\affiliation{CeFEMA-LaPMET, Departamento de Física, Instituto Superior Técnico,
Universidade de Lisboa, Av. Rovisco Pais, 1049-001 Lisboa, Portugal}
\affiliation{Centro de Física das Universidades do Minho e Porto, Departamento de Física e Astronomia, Faculdade de Ciências, Universidade do Porto, 4169-007 Porto, Portugal}
\author{Pedro Ribeiro}
\affiliation{CeFEMA-LaPMET, Departamento de Física, Instituto Superior Técnico,
Universidade de Lisboa, Av. Rovisco Pais, 1049-001 Lisboa, Portugal}
\affiliation{Beijing Computational Science Research Center, Beijing 100193, China}
\author{Ivan M. Khaymovich}
\affiliation{Nordita, Stockholm University and KTH Royal Institute of Technology
Hannes Alfvéns väg 12, SE-106 91 Stockholm, Sweden}
\affiliation{Institute for Physics of Microstructures, Russian Academy of Sciences, 603950 Nizhny Novgorod, GSP-105, Russia}
\begin{abstract}
Quasiperiodic systems in one dimension can host non-ergodic states, e.g. localized in position or momentum. Periodic quenches within localized phases yield Floquet eigenstates of the same nature, i.e. spatially localized or ballistic. However, periodic quenches across these two non-ergodic phases were thought to produce ergodic diffusive-like states even for non-interacting particles. 
We show that this expectation is not met at the thermodynamic limit where the system always attains a non-ergodic state. 
We find that ergodicity may be recovered by scaling the Floquet quenching period with system size and determine the corresponding scaling function. 
Our results suggest that while the fraction of spatially localized or ballistic states depends on the model's details, all Floquet eigenstates belong to one of these non-ergodic categories. Our findings demonstrate that quasiperiodicity hinders ergodicity and thermalization, even in driven systems where these phenomena are commonly expected.
\end{abstract}

\maketitle
The study of localization and ergodicity in quantum many-body systems
has long been a prominent topic of research in Condensed Matter Physics.
Among these studies, the existence of many-body localization (MBL)
and transitions between ergodic and MBL phases in interacting systems
is a hot topic, that is currently under intense scrutiny \citep{PhysRevE.102.062144,PhysRevLett.124.243601,ABANIN2021168415,PhysRevLett.119.075702,PhysRevB.96.104205,PhysRevB.105.224203,PhysRevB.107.115132}.
A different research direction, that dates back to the paradigmatic
Anderson localization \citep{Anderson1958}, focuses in the non-interacting
limit, where nontrivial localization properties can already occur
and a considerably higher degree of understanding can be attained.
Currently, the non-interacting limit is not only of fundamental theoretical
interest, but also very relevant experimentally, since it can be simulated
in optical lattices, where interactions can be tuned \citep{Bloch2005}.
While in the absence of interactions, any finite amount of random
disorder localizes the wave function in 1D short-range Hamiltonians
\citep{Abrahams1979,MacKinnon1981}, non-ergodic ballistic, localized
and even multifractal phases can occur in 1D quasiperiodic systems\citep{PhysRevLett.104.070601,PhysRevLett.113.236403,Liu2015,PhysRevB.91.235134,PhysRevLett.114.146601,anomScipost,GoncalvesRG2022,Devakul2017,critical_phase_goncalves,PhysRevB.93.104504,AubryAndre,PhysRevB.106.205119,PhysRevB.107.014204}, in various long-range models~\cite{Kravtsov_2015,Biroli_RP,PhysRevB.99.104203,Nosov2019mixtures,Motamarri2021RDM,Tang2022nonergodic}, as well as claimed in some hierarical graphs~\cite{RRGAnnals,Altshuler2016nonergodic,Kravtsov2018nonergodic,Tikhonov2016fractality,Tikhonov2017multifractality,Tikhonov2021from}.
A simple but non-trivial paradigmatic model where such physics can
be well understood, is the Aubry-André model, for which an energy-independent
ballistic-to-localized transition occurs at a finite strength of the
quasiperiodic potential \citep{AubryAndre,Roati2008}. 

While the study of localization and ergodicity in periodically driven
systems dates back to the periodically kicked quantum rotator \citep{PhysRevLett.49.509,PhysRevLett.73.2974,PhysRevA.29.1639},
it has experienced a resurgence of interest \citep{PONTE2015196,PhysRevLett.114.140401,PhysRevLett.115.030402,ABANIN20161,PhysRevB.94.020201,PhysRevB.94.094201,PhysRevA.98.053631,PhysRevX.4.041048,Russomanno_2015,PhysRevB.93.104203,PhysRevE.90.052105,PhysRevE.93.012130,PhysRevLett.116.250401}
due to the possibility to emulate time-periodic Hamiltonians and quasiperiodic
potentials in experiments involving ultracold atoms and trapped ions
experiments \citep{RevModPhys.89.011004,Bordia2017,shimasaki2022anomalous}.
These (non-equilibrium) Floquet systems are very appealing, because
on the one hand, they provide a means to realize complex effective
time-independent Hamiltonians by careful choice of the driving
protocol \citep{PhysRevLett.108.225304,PhysRevX.4.031027,Struck2013,Jamotte2022,PhysRevLett.111.185301,PhysRevLett.111.185302,Jotzu2014,zhanwu2016,Aidelsburger2015},
and, on the other hand, they can support novel phases of matter with no
equilibrium counterpart \citep{PhysRevX.6.041001,Yates2022,Kitamura2022,PhysRevLett.106.220402,PhysRevB.82.235114,PhysRevX.3.031005}.
A notable example of the latter arises in interacting 1D quasiperiodic
systems, where driving can induce a transition from non-ergodic many-body-localized
states to ergodic states \citep{PhysRevX.4.041048,PhysRevLett.115.030402,ABANIN20161,Bordia2017,PhysRevA.98.053631}. 

For 1D quasiperiodic systems, the localization phase diagram of the
Floquet Hamiltonian can show a complex structure at high frequencies,
even in the non-interacting limit. Non-ergodic ballistic, localized
and multifractal phases and energy-dependent transitions between them
can arise in the Floquet Hamiltonian \citep{PhysRevLett.87.066601,PhysRevB.90.054303,PhysRevA.98.013635,PhysRevB.99.014301,10.21468/SciPostPhys.4.5.025,PhysRevB.106.054312,PhysRevB.103.184309}, even if they are not present
in the undriven model.
Interestingly, one of the widely studied non-interacting models was
recently realized experimentally in cold atoms \citep{shimasaki2022anomalous}.
For lower frequencies, transitions into a non-ergodic delocalized
phase were reported theoretically even in the absence of interactions
\citep{PhysRevE.97.010101,Romito2018}, where a connection with the
frequency-induced ergodic-to-MBL transition observed experimentally
in Ref.$\,$\citep{Bordia2017} was made. However, these theoretical
studies were mostly carried out for fixed system sizes, possibly motivated
by the limited sizes in cold atom experiments. It is however of paramount
importance to understand the nature of the thermodynamic-limit state,
which requires a detailed finite-size scaling analysis. 

In this paper we carry out a finite-size scaling analysis at large
driving periods for a periodically-driven Aubry-André model and show
that, contrary to previous expectations \citep{PhysRevE.97.010101,Romito2018},
quenches between ballistic and localized states yield non-ergodic
Floquet states in the thermodynamic limit for any finite driving period.
We find that quenches between localized states yield localized Floquet
states as expected \citep{PhysRevE.93.062205,PhysRevB.96.014201,PhysRevLett.123.266601,PhysRevE.97.010101},
and quenches between (non-ergodic) ballistic states yield ballistic
states. However, for quenches between ballistic and localized states,
while ergodic states can be observed for fixed system sizes in the
limit of a large driving period, they flow either to non-ergodic ballistic
or localized states as the system size increases, with fractions that
depend on the center of mass of the quench. 

\section{Model and methods}

\label{sec:Model_Methods}

We consider a periodically-driven Aubry-André model \citep{AubryAndre,PhysRevE.97.010101},
a tight-binding chain of spinless fermions with nearest-neighbor hoppings
and with time-periodic quenches in the quasiperiodic potential. The
Hamiltonian reads
\begin{eqnarray}
H(t)&= & -J\sum_{n=0}^{L-1}   c_{n}^{\dagger}c_{n+1} +\textrm{h.c.} \label{eq:Hamiltonian} \nonumber   \\
& & +\sum_{n}V[1+\epsilon f(t)]\cos(2\pi\tau n+\phi)c_{n}^{\dagger}c_{n} 
\end{eqnarray}
where $c_{n}^{\dagger}$ creates a particle at site $n$, and $L$ is the number of sites/system size. $J$ is the nearest-neighbor hoppings amplitude. We consider twisted boundary conditions, i.e. $c_L = c_0 e^{i\kappa}$, with phase twist $\kappa$.
The last term contains a time-dependent
quasiperiodic modulation of strength $V[1+\epsilon f(t)]$, where
$f(t)\equiv\Theta(t-T/2)-\Theta(T/2-t)$ implements quenches of period
$T$ between Hamiltonians $H_{\pm}$ with quasiperiodic potentials
of strength $V[1\pm\epsilon]$. 

Henceforth, we refer to $V$ as the
center of mass of the quench and set $J=1$. For $\epsilon=0$, the Hamiltonian loses
its time-dependence and we recover the static Aubry-André model, for
which there is a ballistic (localized) phase for $|V|<|J|$ ($|V|>|J|$)
\citep{AubryAndre}.

Throughout the paper, we take $\tau=2/(\sqrt{5}-1)$ (inverse of the golden ratio) in the numerical calculations. To avoid
boundary defects, we consider rational approximations of $\tau$,
$\tau_{c}^{(n)}=F_{n-1}/F_{n}$, for each system size $L_{n}=F_{n}$,
where $F_{n}$ is the $n$-th Fibonacci number \citep{PhysRevLett.43.1954,PhysRevLett.51.1198}.
This choice ensures that the system has a single unit cell for any
system size, being therefore incommensurate. We will consider system
sizes corresponding to Fibonacci numbers in the range $L\in[377-6765]$
($[F_{14}-F_{20}]$). Finally, we also average all the results over
random configurations of the phase twist $\kappa$ and the phase $\phi$
of the quasiperiodic potential. 

The time-evolution operator $U(T)$ for the periodic quench can be
written as 

\begin{equation}
U(T)=e^{-i\int_{0}^{T}H(t)\textrm{ }dt}=e^{-iH_{-}T/2}e^{-iH_{+}T/2}\equiv e^{-iH_{F}T},
\label{U_T}
\end{equation}
where in the last equality we defined the Floquet Hamiltonian $H_{F}$.
The eigenvalues and eigenstates of this Hamiltonian correspond respectively
to the Floquet quasienergies $E_{\alpha}$ and eigenstates
$\ket{\bm{\psi}^{\alpha}}=\sum_{n}\psi_{n}^{\alpha}\ket n$, that
we will study throughout this paper. 

To study the ergodicity of the eigenstates we analysed the energy
level statistics of quasienergies \citep{haake2001quantum,PhysRevB.55.1142},
while to study their localization properties, we computed inverse
participation ratios for the Floquet eigenstates \citep{Janssen,RevModPhys.80.1355}.

For the level statistics analysis, we first order the quasienergies
$\{E_{\alpha}\}$ in the interval $]-\pi,\pi]$ and then compute
consecutive spacings between them, $s_{\alpha}=E_{\alpha+1}-E_{\alpha}$.
We will study the distribution of ratios $r_{\alpha}$ defined as
\citep{PhysRevB.75.155111,PhysRevLett.110.084101},

\begin{equation}
r_{\alpha}=\frac{{\rm min}(s_{\alpha},s_{\alpha-1})}{{\rm max}(s_{\alpha},s_{\alpha-1})}.
\end{equation}

Non-ergodic energy levels are expected to show Poisson (or even sub-Poisson)
statistics, following a distribution $P(r)=2/(1+r)^{2}$, with $\langle r\rangle\approx0.386$,
while ergodic energy levels show level repulsion, following the Gaussian
unitary ensemble (GUE) distribution with $\langle r\rangle\approx0.6$
for systems belonging to the unitary class (which is the case for
our model in Eq.~\eqref{eq:Hamiltonian}, that breaks time-reversal
symmetry due to the phase twists). We note that in fact, since $E_{\alpha}$
are phases, they should follow circular ensembles in the ergodic cases
\citep{PhysRevX.4.041048}. Nonetheless, the distributions obtained
for the circular ensembles should coincide with the distributions
of the corresponding Gaussian ensembles in the thermodynamic limit
\citep{PhysRevX.4.041048}.

For the localization analysis, we computed the real- and momentum-space
inverse participation ratios, given respectively for the Floquet eigenstate
$\ket{\psi^{\alpha}}$ by:

\begin{equation}
\begin{aligned}{\rm IPR}^{\alpha}=\Bigl(\sum_{n}|\psi_{n}^{\alpha}|^{2}\Bigr)^{-2}\sum_{n}|\psi_{n}^{\alpha}|^{4}\propto L^{-D_{r}}\\
{\rm IPR}_{K}^{\alpha}=\Bigl(\sum_{k}|\Phi_{k}^{\alpha}|^{2}\Bigr)^{-2}\sum_{k}|\Phi_{k}^{\alpha}|^{4}\propto L^{-D_{k}}
\end{aligned}
,
\end{equation}
where $\Phi_{k}^{\alpha}=L^{-d/2}\sum_{n}e^{-2\pi\textrm{i}nk}\psi_{n}^{\alpha}$ is the momentum-space wavefunction, and $d=1$  the system's dimension. Since our model does not have any other special basis, other than the real- and momentum-space ones, we focus only on the above two IPRs. States with different localization properties can therefore be distinguished by these quantities: (i) ballistic:
$D_{r}=d$ and $D_{k}=0$; (ii) localized: $D_{r}=0$ and $D_{k}=d$;
fractal/multifractal: $0<D_{r},D_{K}<d$; \footnote{Strictly speaking we can have fractality only in real-space, implying that
$0<D_{r}<D_{K}=d$,
or in momentum-space, implying
$0<D_{K}<D_{r}=d$, see, e.g.,~\cite{PhysRevB.99.104203,Nosov2019mixtures}. However, this is not typically the case for 1D quasiperiodic systems. } (iii) diffusive: $D_{r}=D_{k}=d$.
Henceforth, we define the inverse participation ratios averaged (geometrically) over
all eigenstates and configurations of $\phi$ and $\kappa$ as $\mathcal{I}_{r}=\langle{\rm IPR}^{\alpha}\rangle_{\phi,\kappa,\alpha}$
and $\mathcal{I}_{k}=\langle{\rm IPR}_{K}^{\alpha}\rangle_{\phi,\kappa,\alpha}$.
We averaged over a number of configurations in the interval $N_{c}\in[400,10^{4}]$,
choosing the larger numbers of configurations for the smaller system
sizes.

\section{Results }

\begin{figure}[h]
\centering{}\includegraphics[width=1\columnwidth]{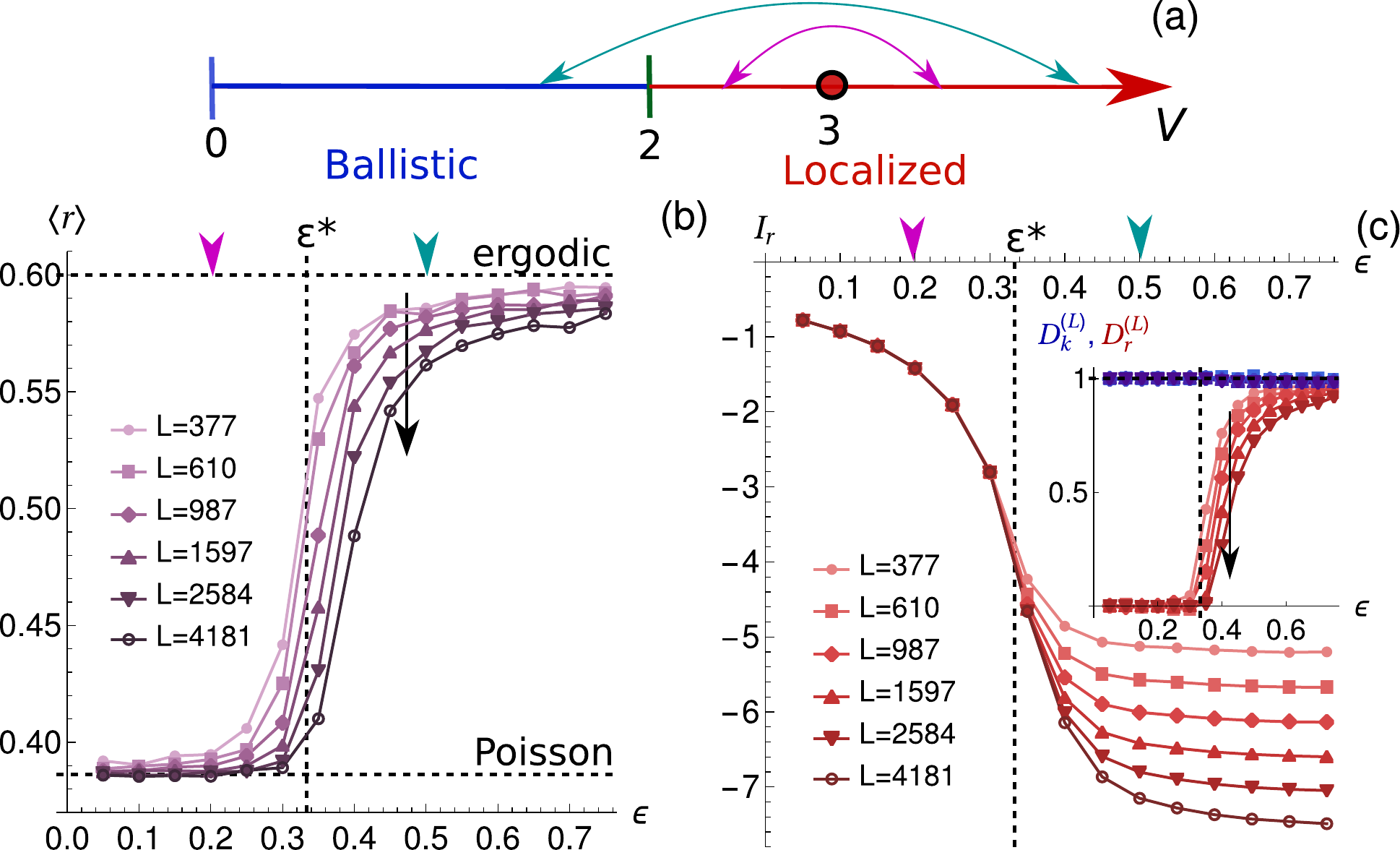}\caption{(a) Floquet-drive location on the static phase diagram  of the Aubry-André model. The arrows indicate
two amplitudes $\epsilon$ of quasiperiodic potentials used in the periodic
quench for $V=3$, also shown in (b,c):
quench between localized states (magenta) and between ballistic
and localized states (cyan). (b) $\langle r\rangle$ for $V=3$ and
$T=500$, as a function of $\epsilon$, for different system sizes
$L$. The vertical dashed line indicates the value $\epsilon=\epsilon^{*}$
above which ballistic and localized states start being quenched, while
the horizontal dashed lines indicate $r_{{\rm GUE}}$ (ergodic) and
$r_{{\rm Poisson}}$. The arrow points towards larger $L$. (c) $\mathcal{I}_{r}$
for the same parameters as in (b). The inset contains the real- and
momentum-space finite-size fractal dimensions defined in Eq.$\,$\eqref{eq:fractal_dims}.
\label{fig:1}}
\end{figure}

\textbf{\textit{Quench's center of mass at localized phase.---}}
We start by studying the case where the center-of-mass of the quench
lies in the localized phase. For this purpose, we set $V=3$, a large
driving period $T=500$ and vary the Floquet amplitude $\epsilon$. In Fig.~\ref{fig:1},
we show the results for $\langle r\rangle$ and $I_{r}$, averaged
over all the Floquet eigenstates. In the inset of Fig.~\ref{fig:1}(c),
we also show the finite-size fractal dimensions $D_{r}^{(L)}\equiv\mathcal{D}^{(L)}(\mathcal{I}_{r})$
and $D_{k}^{(L)}\equiv\mathcal{D}^{(L)}(\mathcal{I}_{k})$, where 

\begin{equation}
\mathcal{D}^{(L=L_{n})}(\mathcal{I})=\frac{\log[\mathcal{I}(L_{n+1})]-\log[\mathcal{I}(L_{n})]}{\log(L_{n+1})-\log(L_{n})},\label{eq:fractal_dims}
\end{equation}

\begin{figure}[H]
\centering{}\includegraphics[width=1\columnwidth]{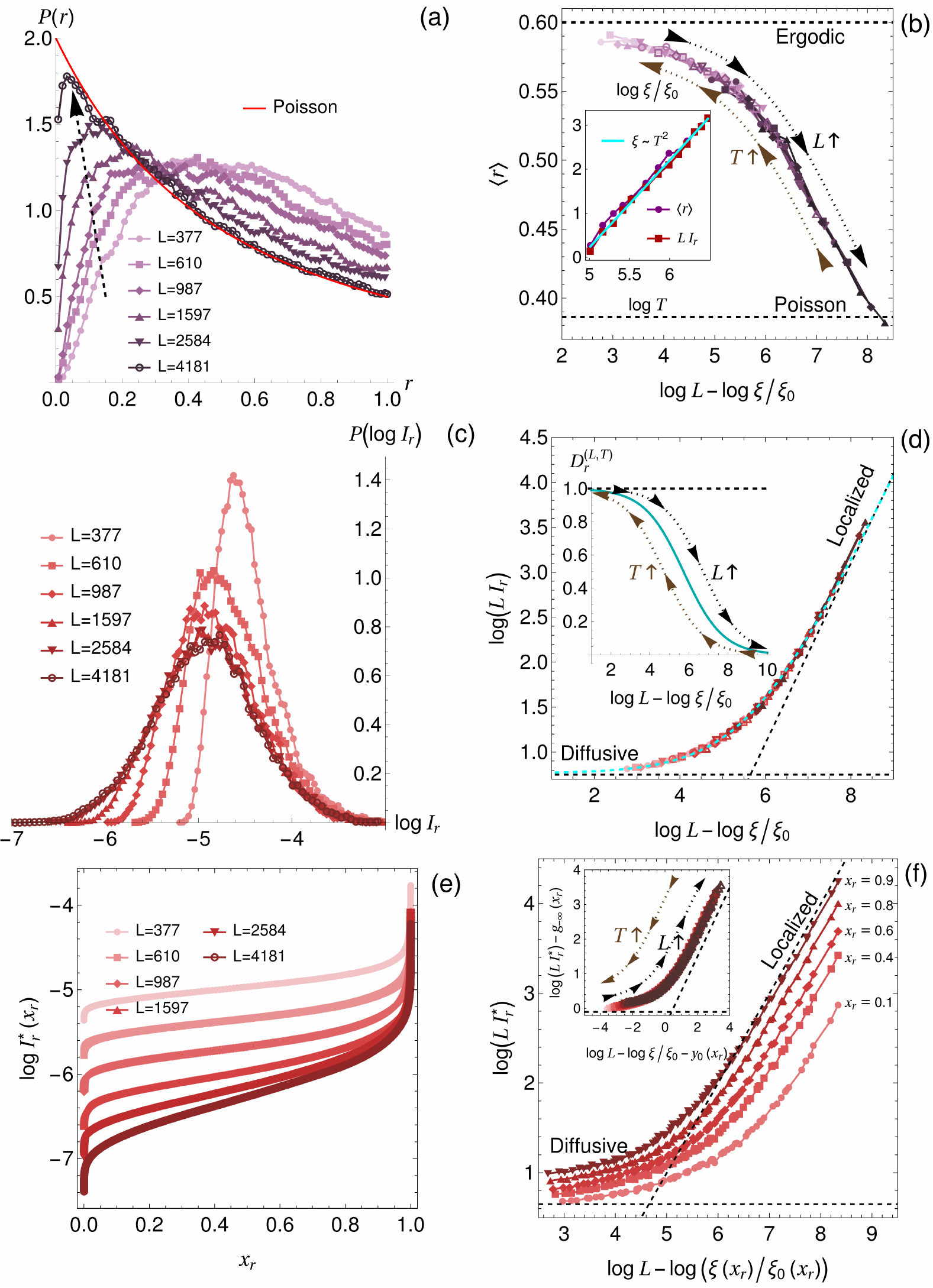}\caption{\textbf{Quench's center of mass in localized phase: }results for $V=3,\epsilon=0.45> \epsilon^*$.
(a) $P(r)$ distribution at $T=150$ for different system sizes.
The arrow points towards increasing $L$ and the full red line indicates
the Poisson distribution $P(r)=2/(1+r)^{2}$. The distributions were
computed by binning, where each data point corresponds to the bin
center. (b) Scaling collapse of $\langle r\rangle$ calculated for
different driving periods $T$ belonging to the interval $T\in[125,650]$
and $L\in[377,4181]$ (see Appendix for details on the scaling collapse).
The correlation lengths $\xi$ were extracted from the scaling collapse
and $\xi_{0}=\xi(T=125)$. The black (brown) arrows point towards
increasing $L$ ($T$). The inset shows $\log(\xi/\xi_{0})$ as function
of $\log T$, obtained from both the collapses of $\langle r\rangle$
and of $L\mathcal{I}_{r}$ done in (d). The cyan line was obtained
by fitting for the data points at $6$ largest $T$-values (combining the
data from $\langle r\rangle$ and $L\mathcal{I}_{r}$ ), yielding
a power-law $\xi\sim T^{2}$. (c) Distribution $P(\log\mathcal{I}_{r})$
for $T=150$ and different system sizes. (d) Scaling collapse of
$L\mathcal{I}_{r}$, again for $T\in[125,650]$. The large $L$ localized
behaviour $L\mathcal{I}_{r}\sim L$ and large $\xi$ (or $T$) diffusive
behaviour $L\mathcal{I}_{r}\sim L^{0}$ are indicated by the black
dashed lines. We find that $\log(L\mathcal{I}_{r})$ is well fitted
by the expression $g(y)=g_{-\infty}+\log(1+e^{y-y_{0}})$, where $y=\log L-\log\xi/\xi_{0}$,
shown in the dashed cyan line. We obtain the fractal dimension from
this fit through $D_{r}^{(L,T)}=1-[\log(L\mathcal{I}_{r})]'$, and
show in the inset, in full cyan. (e) $\log\mathcal{I}_{r}^{*}(x_{r})$
defined in Eq.~\eqref{eq:xr} and below it, for different $L$ and
for $T=300$. (f) Scaling collapse of $\log\mathcal{I}_{r}^{*}(x_{r})$
for different values of $x_{r}$ indicated in the figure. We note
that the computed $\xi(x_{r})$ are identical for the different $x_{r}$,
as shown in Appendix \ref{sec:scaling_collapses}. By fitting the
collapsed data for different $x_{r}$ to $g^{(x_{r})}(y)=g_{-\infty}(x_{r})+\log(1+e^{y-y_{0}(x_{r})})$
we were able to collapse all the curves into an universal curve shown
in the inset. The fitted parameters $g_{-\infty}(x_{r})$ and $y_{0}(x_{r})$
are shown in Fig.~\ref{fig:y0_g-inf}(a) of the Appendix. \label{fig:2}}
\end{figure}

\noindent such that $D_{p}^{(\infty)}=D_{p},\textrm{ }p=r,k$. When $\epsilon<\epsilon^{*}=1-2/V(V>2)$,
the quench is only between localized states, as illustrated in Fig.~\ref{fig:1}(a).
In this case, we see that up to weak finite-size effects, $\langle r\rangle=r_{\textrm{Poisson}}$,
$D_r=0$ and $D_{k}=1$, clearly showing that the Floquet eigenstates
are non-ergodic and localized. Once $\epsilon\geq\epsilon^{*}$, we
start quenching between ballistic and localized phases, which is accompanied
by a sharp increase in $\langle r\rangle$, that approaches $r_{\textrm{GUE}}$
as $\epsilon$ is increased; and in $D_r^{(L)}$, that approaches $1$,
while we still have $D_{k}^{(L)}\approx1$. Upon initial observation,
this behaviour could indicate a transition into a diffusive ergodic
phase. However, when $L$ is increased, there is a clear overall decrease
both in $\langle r\rangle$ and in $D$, which is already a clear
indication of the fragile nature of the ergodic-phase candidate.

The instability of the ergodic phase is further corroborated in Fig.~\ref{fig:2},
where we also set $V=3$ and choose $\epsilon=0.45>\epsilon^{*}$,
to quench between ballistic and localized states. In Fig.~\ref{fig:2}(a)
we show the distribution of ratios $P(r)$ for fixed $T=150$ and
for different system sizes. There, we can clearly see that the distribution
of ratios transitions from exhibiting level repulsion to closely resembling
the Poisson distribution as $L$ is increased. Concurrently, Fig.~\ref{fig:2}(c)
demonstrates that the distribution of $\log\mathcal{I}_{r}$ for the
same $T$ is almost entirely converged for the larger $L$ used, implying
the localization of all (or nearly all) states {[}note that $D_{k}\approx1$,
as shown in Fig.~\ref{fig:1}(c){]}. 

The results so far are in support of an ergodic phase only surviving
when $T\rightarrow\infty$ at finite $L$. With this in mind, we define
a correlation length $\xi(T)$ that diverges when $T\rightarrow\infty$,
such that when $L\ll\xi$, the system is ergodic while when $L\gg\xi$,
the system is non-ergodic. Close to the transition to the diffusive
ergodic phase, that is, for large enough $T$, we assume that $\xi$
diverges as a power-law in $T$, $\xi\sim T^{\beta}$, with an unknown
exponent $\beta$ that may depend on the model parameters. We also
assume that in this regime, $\langle r\rangle$ follows a one-parameter
scaling function that satisfies,

\begin{equation}
\langle r\rangle=f(L/\xi)=\begin{cases}
r_{\textrm{GUE}} & ,L\ll\xi\\
\approx r_{\textrm{Poisson}} & ,L\gg\xi
\end{cases}.\label{eq:r_scaling_func}
\end{equation}

In a similar way, we also assume $\mathcal{I}_{r}$ follows the one-parameter
scaling ansatz $\mathcal{I}_{r}=L^{\mu}g(L/\xi)$ at large enough
$T$. Using that for $\xi/L\rightarrow\infty$, $\mathcal{I}_{r}=L^{\mu}g(0)\sim L^{-1}$
(diffusive and ergodic), we get that $\mu=-1$. In the limit $L/\xi\rightarrow\infty$,
the states are localized, $\mathcal{I}_{r}\sim L^{0}$.
We therefore have the following limits for $g(L/\xi)$
\begin{equation}
L\mathcal{I}_{r}=g(L/\xi)\sim \begin{cases}
1 & ,L\ll\xi\\
L/\xi & ,L\gg\xi 
\end{cases} . \label{eq:Ir_scaling_func}
\end{equation}

As a consequence $\mathcal{I}_{r} \simeq 1/\min(L, \xi)$, meaning that $\xi(T)$ is, indeed, a $T$-dependent localization length of the model. Indeed, as soon as $L\ll \xi$, the states do not know about $\xi$ and look like ergodic ones, while in the opposite limit of $L\gg \xi$, the boundary conditions are not important and all the states are localized at a distance $\sim \xi$.

We note that here we are not considering the scaling function for
$\mathcal{I}_{k}$, since $\mathcal{I}_{k}\sim L^{-1}$ both in the
diffusive and localized phases. 

 In Figs.~\ref{fig:2}(b,d), we collapse data for different periods
in the range $T\in[125,650]$ and for different $L$, showing the
validity of the scaling ansatzes in Eqs.~\eqref{eq:r_scaling_func},~\eqref{eq:Ir_scaling_func}.
In Appendix~\ref{sec:scaling_collapses} we provide precise details
on how the scaling collapses were computed. From the collapses, we
can extract $\xi(T)$ that we plot in the inset of Fig.~\ref{fig:2}(b).
In this figure, we see that $\xi(T)$ acquires a power-law behaviour
at large $T$, as expected, giving compatible results when extracted
from the scaling collapses of $\langle r\rangle$ and $\mathcal{I}_{r}$.
By fitting the power-law at large $T$, we extract $\beta=2$. We
note however, that this exponent is non-universal and depends on the
model's parameters as we demonstrante below for other examples. 

The good scaling collapses in Figs.~\ref{fig:2}(b,d) confirm
our previous affirmations: (i) when the system size increases for
fixed $T$ (that is, fixed $\xi$), the Floquet eigenstates flow to
a non-ergodic localized phase; (ii) if $T$ is increased for fixed
$L$, the system flows to a diffusive ergodic phase. This implies
that the limits $T\rightarrow\infty$ and $L\rightarrow\infty$ do
not commute.

Next, in order to inspect how different parts of the $\mathcal{I}_{r}$
distribution evolve with $T$ and $L$, similarly to~\cite{10.21468/SciPostPhys.4.5.025}, we define the fraction of
states $x_{r}$ for which the average IPR is bounded by $\mathcal{I}_{r}=\mathcal{I}_{r}^{*}$,
given by

\begin{equation}
x_{r}=\int_{-\infty}^{\log\mathcal{I}_{r}^{*}}P(y')dy'\label{eq:xr}
\end{equation}
where $y'=\log\mathcal{I}_{r}$. In Fig.~\ref{fig:2}(e) we plot
$\log\mathcal{I}_{r}^{*}(x_{r})$, showing that it is a smooth function
of $x_{r}$. To analyse how $\mathcal{I}_{r}^{*}$ evolves for different
fractions $x_{r}$, in Fig.~\ref{fig:2}(f) we perform the $x_r$-dependent collapse of $\log L\mathcal{I}_{r}^{*}(x_{r})$. We observe that the corresponding
scaling functions have the properties of Eq.~\eqref{eq:Ir_scaling_func}
for the studied fractions of states $x_{r}$ and can even be collapsed
into a single universal curve given in the inset of Fig.~\ref{fig:2}(f),
as detailed in the Figure's caption. Noteworthy, we checked that the
correlation lengths $\xi(x_{r})$ obtained from the scaling collapses
at different $x_{r}$ are almost independent of $x_{r}$ at large
$T$, as we show in Appendix \ref{sec:scaling_collapses}.

In Appendix \ref{sec:ballistic_COM}, we also studied the case when
the quench's center of mass lies in the ballistic phase. In this case,
we observed that the Floquet eigenstates also become non-ergodic in
the thermodynamic limit, but ballistic, instead of localized. 

\begin{figure}[h]
\centering{}\includegraphics[width=1\columnwidth]{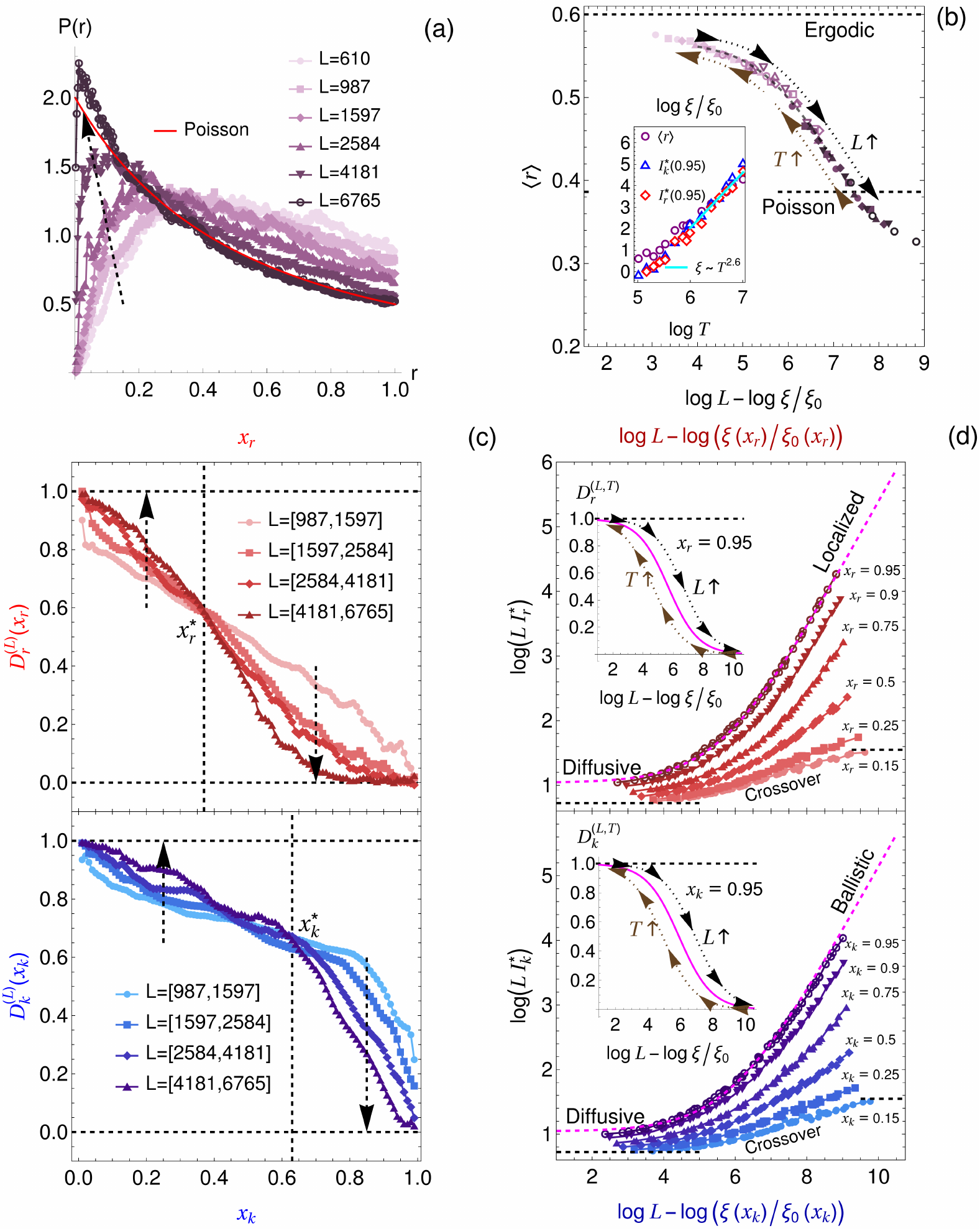}\caption{\textbf{Quench's center of mass at critical point: }results for $V=2,\epsilon=0.2$.
(a) $P(r)$ distribution for $T=250$, for different system sizes
$L$. (b) Scaling collapses of $\langle r\rangle$ calculated for
driving periods $T$ belonging to the interval $T\in[125,1100]$ and
for $L\in[1597-6765]$ (see Appendix for details on the scaling collapse).
The inset shows $\log(\xi/\xi_{0})$ as function of $\log T$, obtained
from the data collapses of $\langle r\rangle$, $L\mathcal{I}_{r}^{*}(x_{r}=0.95)$
and $L\mathcal{I}_{k}^{*}(x_{k}=0.95)$ shown in (d). The cyan line
was obtained by fitting for the data points at $7$ largest $T$-values
(combining the data from the three shown collapses), yielding a power-law
$\xi\sim T^{2.6}$. (c) Fractal dimensions $D_{r}^{(L)}(x_{r})\equiv\mathcal{D}\big(\mathcal{I}_{r}^{*}(x_{r})\big)$
(top) and $D_{k}^{(L)}(x_{k})\equiv\mathcal{D}\big(\mathcal{I}_{k}^{*}(x_{k})\big)$
(bottom) {[}see Eq.~\eqref{eq:fractal_dims}{]}, for $T=125$. (d)
Scaling collapses for $\log\big(L\mathcal{I}_{r}^{*}(x_{r})\big)$
(top) and $\log\big(L\mathcal{I}_{k}^{*}(x_{k})\big)$ (bottom), for
different $x_{r}$ and the same range of $T$ and $L$ as in (a).
The magenta dashed lines correspond to fits of the collapsed data
for $x_{r}=0.95$ (top) and $x_{k}=0.95$ (bottom) to the function
$g(y)=g_{-\infty}+\log(1+e^{y-y_{0}})$, with $y=\log L-\log\xi/\xi_{0}$.
In the inset, we show the $L$- and $T$-dependent fractal dimensions
obtained from these fits through $D_{r(k)}^{(L,T)}\equiv1-[\log(L\mathcal{I}_{r(k)})]'$.\label{fig:3}}
\end{figure}

\textbf{\textit{Quench's center of mass at critical point.---}} We
now turn to the case where the quench's center of mass is exactly
at the critical point, that is, $V=2$. This is studied in Fig.~\ref{fig:3}.
Similarly to the previous case, Fig.~\ref{fig:3}(a) reveals the
evolution breakdown of level repulsion when $L$ increases,
for fixed $T$. In Fig.~\ref{fig:3}(b) we see that a scaling
collapse for $\langle r\rangle$ is still possible. It is worth noticing
that in this case, however, $\langle r\rangle$ can take values significantly
below $r_{\textrm{Poisson}}$ for large $L$. This can however be a finite-size effect arising from the formation of energy gaps that can only be resolved for large enough $L$. In this case, the $P(r)$ distribution should converge to the Poisson distribution in the thermodynamic limit.

The main difference for this quench comparing to the $V=3$ case is
that there is clearly a fraction of states that flow to localized behaviour,
while the remaining fraction flows to ballistic behaviour, as $L$ is increased.
This is illustrated in Fig.~\ref{fig:3}(c). There, we define
$x_{r}$ as in Eq.~\eqref{eq:xr} for $\mathcal{I}_{r}$ and we
also define $x_{k}$ using the analogous definition for $\mathcal{I}_{k}$:

\begin{equation}
x_{k}=\int_{-\infty}^{\log\mathcal{I}_{k}^{*}}P(y')dy'\label{eq:xk}
\end{equation}
where $y'=\log\mathcal{I}_{k}$. For the following discussion, we
define the $x_{p}$-dependent fractal dimensions as $D_{p}^{(L)}(x_{p})=\mathcal{D}\big(\mathcal{I}_{p}^{*}(x_{p})\big)$
(see Eq.~\eqref{eq:fractal_dims}), with $p=r,k$. In Fig.~\ref{fig:3}(c),
we can see that for $x_{r}>x_{r}^{*}\approx0.35$ (see $x_{r}^{*}$
indicated in the figure), $D_{r}^{(L)}(x_{r})$ decreases with $L$,
seemingly towards $0$. Concomitantly, $D_{k}^{(L)}(x_{k})$ increases
towards $1$ for $x_{k}<x_{k}^{*}=1-x_{r}^{*}\approx0.65$. This is
an indication that approximately $65\%$ of states are localized in
the thermodynamic limit. On the other hand, for the remaining fraction
of $\approx35\%$ states, the results are concomitant with $D_{r}^{(L)}(x_{r}<x_{r}^{*})\rightarrow1$
and $D_{k}^{(L)}(x_{k}>x_{k}^{*})\rightarrow0$, as expected for ballistic
states. 

We note that it might happen that a finite fraction of multifractal
states survives in the thermodynamic limit. However, for the available
system sizes, all the states seem to flow to localized and ballistic ones. That being the case, only a fraction
of multifractal states of measure zero, arising at mobility edges
between ballistic and localized states, should survive the thermodynamic
limit. 

Finally, in Fig.~\ref{fig:3}(d) we make scaling collapses of
$\log L\mathcal{I}_{k}^{*}$ and $\log L\mathcal{I}_{r}^{*}$ for
different $x_{r}$ and $x_{k}$. We can see that for large enough
$x_{r}$, $D_{r}$ clearly flows from $D_{r}=1$ (diffusive) to $D_{r}=0$
(localized) as $L\rightarrow\infty$ for fixed $T$. In the same way,
for large enough $x_{k}$, $\Gamma_{k}$ flows from $D_{k}=1$ (diffusive)
to $D_{k}=0$ (ballistic) as $L\rightarrow\infty$ for fixed $T$.
For small $x_{r}$ ($x_{k}$), $D_{r}=1$ ($D_{k}=1$) in both the
limits $L/\xi\rightarrow\infty$ and $\xi/L\rightarrow\infty$, as
indicated by the constant $\log L\mathcal{I}_{r}^{*}$ ($\log L\mathcal{I}_{k}^{*}$)
in both these limits. It is nonetheless interesting to notice that
even in this case there is a crossover regime at finite $L$ and $T$
indicated in Fig.~\ref{fig:3}(d).

\begin{figure}[h]
\centering{}\includegraphics[width=1\columnwidth]{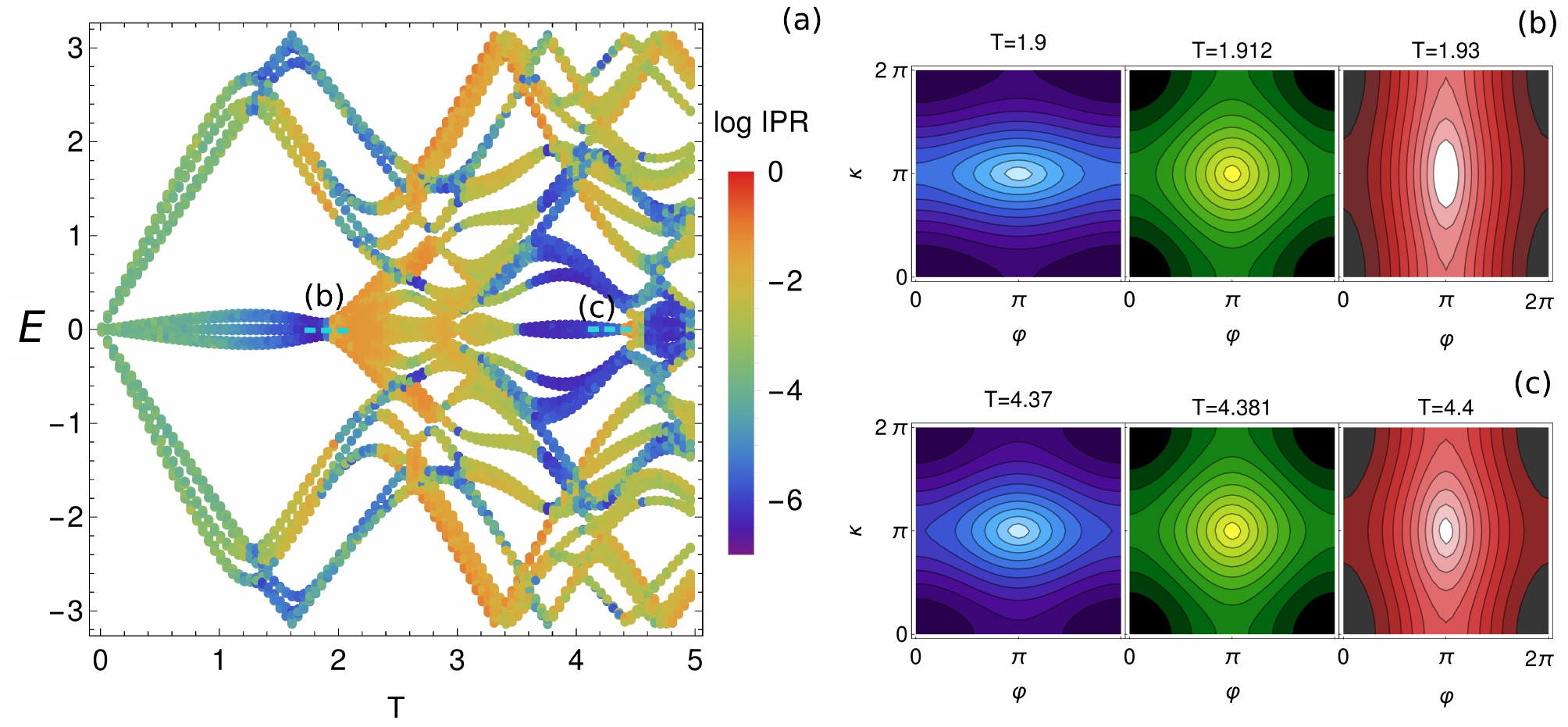} \caption{(a) $\log{\rm IPR}$ as a function of quasienergy $E$ and
driving period $T$, for $L=987,V=2,\epsilon=0.8$, and for a fixed
random choice of $\phi$ and $\kappa$. (b,c) Quasienergy contours
(lighter colors correspond to larger quasienergy) in the plane of
phases $\varphi\equiv L\phi$ and $\kappa$, for $L=34$ and values
of $T$ indicated above each figure, for the $\lfloor L/2\rfloor$-th
largest quasienergy (ordered in the interval $]-\pi,\pi]$), that
has $E\approx0$. The results were obtained around the ballistic-to-localized
transitions indicated by the dashed cyan lines in (a), with the blue
and red figures chosen respectively inside the ballistic and localized
phases, and the green figure approximately at the critical point.
\label{fig:4}}
\end{figure}

\textbf{\textit{Dualities and universality at small $T$.---}} Up
to now, we verified that when $L\rightarrow\infty$, the Floquet eigenstates
become non-ergodic, as in the static limit. At large $T$, however,
there is a very complex structure of mobility edges, and a (quasi)energy-resolved
analysis becomes very challenging. On the other hand, for small $T$,
such analysis is still possible and elucidating. In Fig.~\ref{fig:4}(a),
we show an example where it can be clearly seen that for small $T$,
even though the phase diagram can already be quite complex, clear
transitions between ballistic (low $\textrm{IPR}$) and localized
(large $\textrm{IPR}$) phases can still be found. In the static case,
hidden dualities with universal behaviour were found to be behind
these transitions \citep{HdualitiesScipost}. Moreover, it was found
that ballistic, localized and even critical phases can be understood
in terms of renormalization-group flows to simple renormalized effective
models \citep{GoncalvesRG2022}. 

Remarkably, we find that these results can be generalized for the
Floquet Hamiltonian. This can be seen by inspecting the dependence
of the quasienergies on the potential shift $\varphi\equiv L\phi$
and on the phase twist $\kappa$ \citep{HdualitiesScipost,GoncalvesRG2022}.
We illustrate this for two representative ballistic-to-localized transitions
in Figs.~\ref{fig:4}(b,c), where we see that: (i) at the ballistic
(localized) phase, the quasienergy dependence on $\kappa$ ($\varphi$)
is dominant and the dependence on $\varphi$ ($\kappa$) becomes irrelevant
as $L\rightarrow\infty$ (not shown); (ii) the quasienergies become
invariant under switching $\kappa$ and $\varphi$ at the critical
point. This is exactly the universal behaviour also found for the
single-particle energies in the static case \citep{HdualitiesScipost,GoncalvesRG2022}. 

With these results in mind, we conjecture that the hidden dualities
and RG universality found at small $T$ extend to large $T$, but
only for a large enough system size when the system flows to one of
the non-ergodic phases. It is however very challenging to verify this
conjecture due to the intricate structure of mobility edges at large
$T$ and the limited available system sizes.

\section{Discussion}

Contrary to prior expectations, we have established that time-periodic quenches between non-ergodic ballistic and localized states in non-interacting 1D quasiperiodic systems lead to the emergence of non-ergodic states at the thermodynamic limit, for any finite driving period. To restore ergodicity, the driving period must be scaled with the system size, according to the corresponding scaling functions, which we also determined. 

We expect our findings to hold in generic driven non-interacting 1D quasiperiodic systems. Even though delocalized phases were previously reported for small enough driving frequencies, no clear phase with ergodic properties surviving the thermodynamic limit was identified so far. 

For instance, in Ref.$\,$\citep{PhysRevB.103.184309}, a localization-delocalization transition with decreasing driving frequency was recently reported. However, as we detail in Appendix \ref{loc_deloc_sarkar}, the low-frequency extended phases are non-ergodic, either ballistic or multifractal.

Our findings raise interesting further questions, such as quenching outcomes between distinct phases in higher dimensions, where ergodic states can exist in static, non-interacting situations. These results also suggest that finite interactions may be crucial for the observation of driving-induced ergodic to MBL transitions reported experimentally \citep{Bordia2017}. Nonetheless, it is likely that the ergodic to non-ergodic crossover, which we predict for the non-interacting limit, is experimentally accessible. If so, this would allow the experimental determination of the scaling function between the period and the system size which effectively characterises the fragility of the non-interacting ergodic states.

\section*{Acknowledgments}

\begin{acknowledgments}
M.~G. and P.~R. acknowledge partial support from Fundação para
a Ciência e Tecnologia (FCT-Portugal) through Grant No. UID/CTM/04540/2019.
M.~G. acknowledges further support from FCT-Portugal
through the Grant SFRH/BD/145152/2019. 
I.~M.~K. acknowledges the support
by Russian Science Foundation (Grant No. 21-12-00409).
We finally
acknowledge the Tianhe-2JK cluster at the Beijing Computational Science
Research Center (CSRC), the Bob\textbar Macc supercomputer through
computational project project CPCA/A1/470243/2021 and the OBLIVION
supercomputer, through projects HPCUE/A1/468700/2021, 2022.15834.CPCA.A1
and 2022.15910.CPCA.A1 (based at the High Performance Computing Center
- University of Évora) funded by the ENGAGE SKA Research Infrastructure
(reference POCI-01-0145-FEDER-022217 - COMPETE 2020 and the Foundation
for Science and Technology, Portugal) and by the BigData@UE project
(reference ALT20-03-0246-FEDER-000033 - FEDER and the Alentejo 2020
Regional Operational Program. Computer assistance was provided by
CSRC's, Bob\textbar Macc's and OBLIVION's support teams. 
\end{acknowledgments}



\bibliographystyle{apsrev4-1}
\bibliography{Floquet,Non_interacting_refs}

\newpage
\appendix
\section{Quench's Center of mass deep in ballistic phase}

\label{sec:ballistic_COM}

\begin{figure}[h]
\centering{}\includegraphics[width=1\columnwidth]{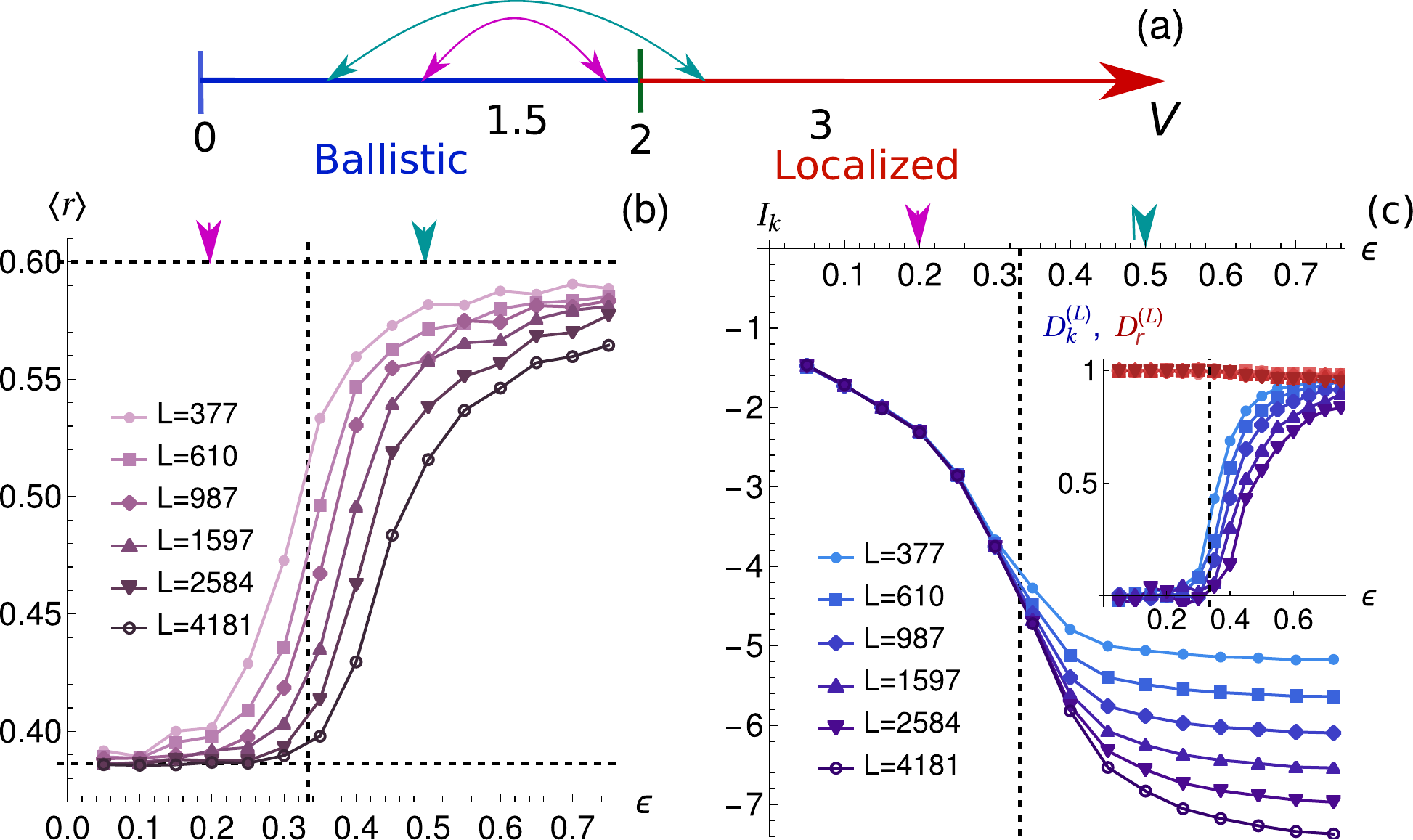}\caption{(a) Floquet-drive location on the static phase diagram  of the Aubry-André model. The arrows indicate
two amplitudes $\epsilon$ of quasiperiodic potentials used in the periodic
quench for $V=1.5$, also shown in (b,c):
quench between ballistic states (magenta); and quench between ballistic
and localized states (cyan). (b) $\langle r\rangle$ for $V=1.5$
and $T=500$, as a function of $\epsilon$, for different system sizes
$L$. The vertical dashed line indicates the value $\epsilon=\epsilon^{*}$
above which we start quenching ballistic and localized states, while
the horizontal dashed lines indicate $r_{{\rm GUE}}$ (ergodic) and
$r_{{\rm Poisson}}$. (c) $\mathcal{I}_{k}$ for the same parameters
as in (b). The inset contains the real- and momentum-space finite-size
fractal dimensions defined in Eq.~\eqref{eq:fractal_dims}. \label{fig:S1}}
\end{figure}

\begin{figure}[H]
\centering{}\includegraphics[width=1\columnwidth]{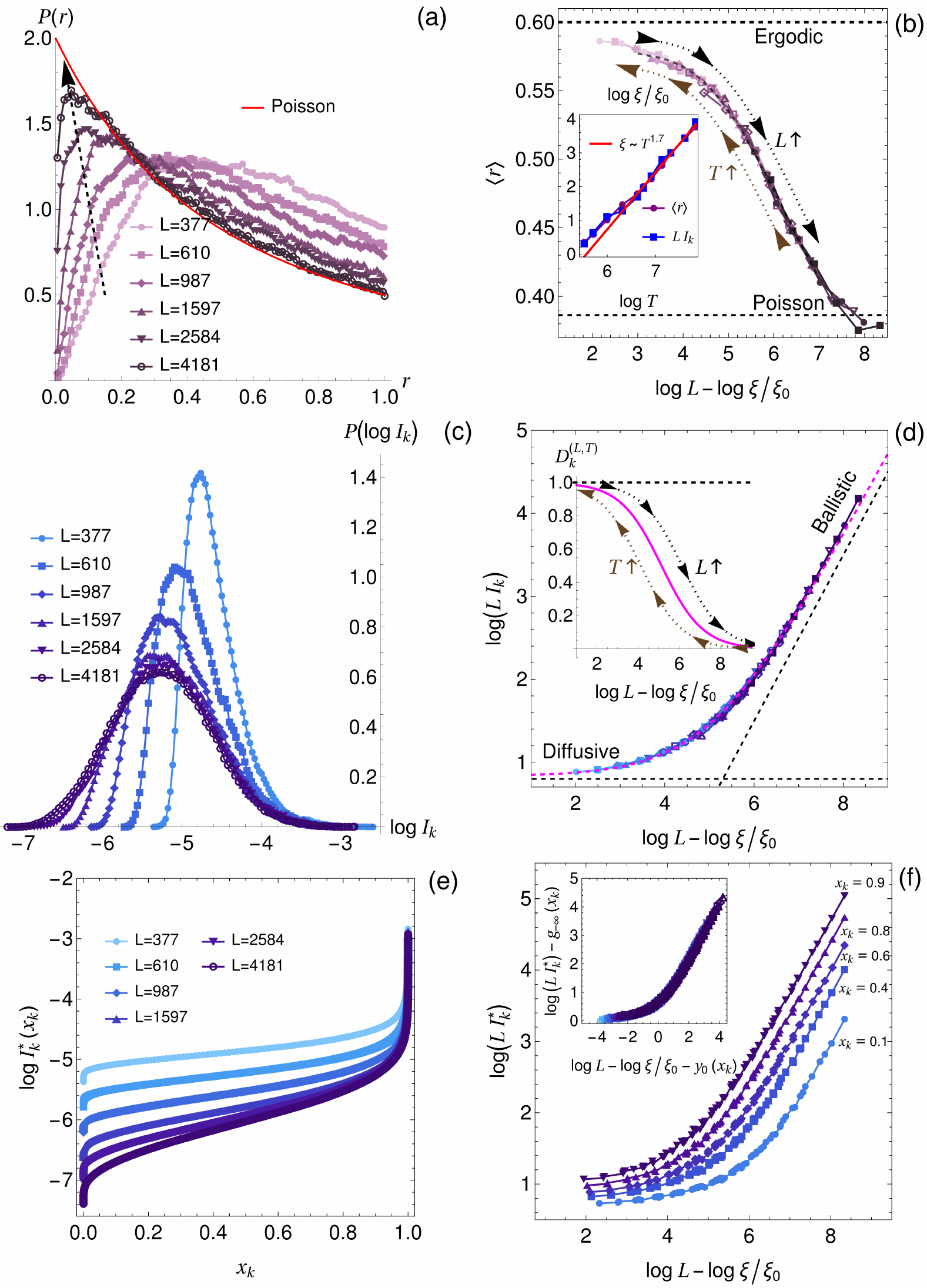}\caption{\textbf{Quench's center of mass in ballistic phase:} Results for $V=1.5,\epsilon=0.375$.
(a) $P(r)$ distribution for $T=400$, for different system sizes.
The arrow points towards increasing $L$ and the full red line indicates
the Poisson distribution $P(r)=2/(1+r)^{2}$. (b) Scaling collapses
of $\langle r\rangle$ calculated for values of $T$ belonging to
the interval $T\in[200,2500]$ and $L\in[377,4181]$. The correlation
lengths $\xi$ were obtained from the scaling collapse and $\xi_{0}=\xi(T=200)$.
The black (brown) arrows point towards increasing $L$ ($T$). The
inset shows $\log(\xi/\xi_{0})$ as function of $\log T$, for both
the collapses of $\langle r\rangle$ and of $L\mathcal{I}_{k}$ done
in (d), yielding compatible results. The cyan line was obtained by
fitting for the data points at $6$ largest $T$-values  (combining the data from
$\langle r\rangle$ and $L\mathcal{I}_{k}$ ), yielding a power-law
$\xi\sim T^{1.7}$. (c) Distribution $P(\log\mathcal{I}_{k})$ for
$T=400$ and different system sizes. (d) Scaling collapses of $L\mathcal{I}_{k}$.
The large $L$ localized behaviour $L\mathcal{I}_{k}\sim L$ and large
$\xi$ (or $T$) diffusive behaviour $L\mathcal{I}_{k}\sim L^{0}$
are indicated by the black dashed lines. We find that $\log(L\mathcal{I}_{k})$
is well fitted by the expression $g(y)=g_{-\infty}+\log(1+e^{y-y_{0}})$,
where $y=\log L-\log\xi/\xi_{0}$, shown in the dashed magenta line.
We obtain the fractal dimension from this fit through $D_{k}^{(L,T)}=1-[\log(L\mathcal{I}_{k})]'$
and show in the inset, in full magenta. (e) $\log\mathcal{I}_{k}^{*}(x_{k})$,
for different $L$ and for $T=850$. (f) Scaling collapses of $\log\mathcal{I}_{k}^{*}(x_{k})$
for different values of $x_{k}$ indicated in the figure. We note
that the computed $\xi(x_{k})$ are identical for the different $x_{k}$,
as shown in Fig.~\ref{fig:xi(x)}(b). By fitting the collapsed
data for different $x_{k}$ to $g^{(x_{k})}(y)=g_{-\infty}(x_{k})+\log\left(1+e^{y-y_{0}(x_{k})}\right)$
we were able to collapse all the curves into an universal curve shown
in the inset. The fitted parameters $g_{-\infty}(x_{k})$ and $y_{0}(x_{k})$
are plotted in Fig.~\ref{fig:y0_g-inf}(b). \label{fig:V1.5.eps-0.375}}
\end{figure}

In this Appendix section, we study a periodic quench with a center-of-mass
in the ballistic phase. In Fig.~\ref{fig:S1}, we see that a quench
that only mixes ballistic states, gives rise to ballistic states,
as signaled by $\langle r\rangle=r_{\textrm{Poisson}},D=1$ and $D_{k}=0$
for $\epsilon<\epsilon^{*}=2/V-1(V<2)$. For $\epsilon\geq\epsilon^{*}$,
we see that even though there is a sudden increase in $\langle r\rangle$
and $D_{k}$ for fixed system sizes towards what is expected in an
ergodic phase (similarly to the increase in $\langle r\rangle$ and
$D$ when the quench was centered at the localized phase, in Fig.~\ref{fig:1}),
the latter behaviour is not robust when $L$ is increased. To prove
this point in a more precise way, we fix $\epsilon=0.375>\epsilon^{*}$
and make a detailed finite-size scaling analysis in Fig.~\ref{fig:V1.5.eps-0.375},
as done in Fig.~\ref{fig:1} for $V=3$. There, we see that all
(or almost all) Floquet eigenstates become ballistic when $L\rightarrow\infty$.

\begin{figure*}
\begin{centering}
\includegraphics[width=1\textwidth]{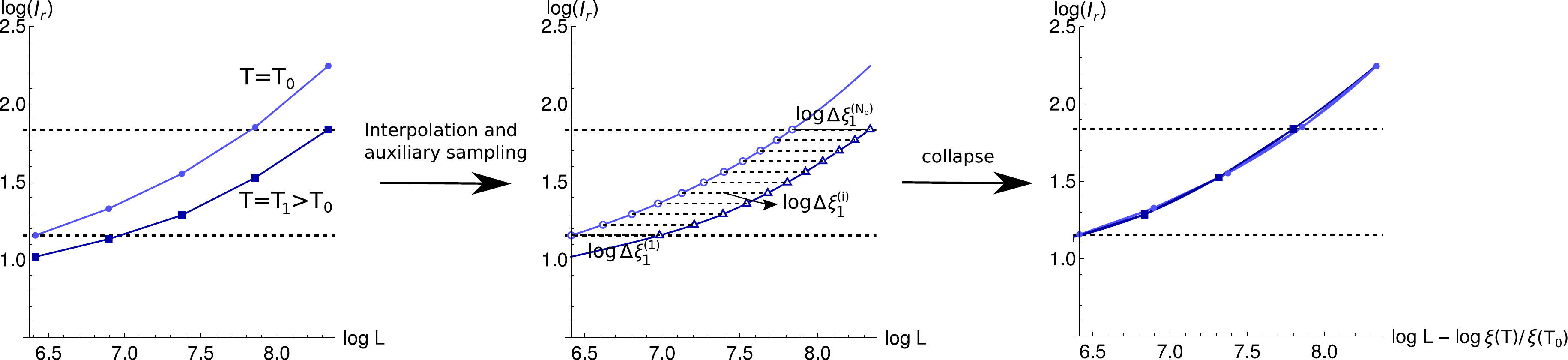}\caption{Method used for data collapse. We first interpolate the original data
points, corresponding to a given $T=T_{0}$ and $T=T_{1}>T_{0}$ and
uniformly sample $N_{p}$ points of the resulting function in the
vertical region of overlap between the two functions (in our calculations
we usually take $N_{p}=10$ and a linear interpolation). We then compute
all the distances between the sampled points $\log\Delta\xi_{1}^{(i)},i=1,\cdots,N_{p}$
and define $\log\Delta\xi_{1}=N_{p}^{-1}\sum_{i}\log\Delta\xi_{1}^{(i)}$.
We then use this distance to collapse the data. Considering $\xi_{0}$
to be the correlation length at $T=T_{0}$, we then have $\log\xi(T_{1})=\log\xi_{0}+\log\Delta\xi_{1}$.
We can then iterate this procedure by taking larger periods $T_{n}$,
always collapsing the curves for $T=T_{n}$ and $T=T_{n-1}$, from
which $\log\Delta\xi_{n}$ can be extracted. We can then compute $\log[\xi(T=T_{n})/\xi_{0}]=\sum_{m=1}^{n}\log\Delta\xi_{m}$.\label{fig:scaling_collapses}}
\par\end{centering}
\end{figure*}

\begin{figure*}[t]
\begin{centering}
\includegraphics[width=1\textwidth]{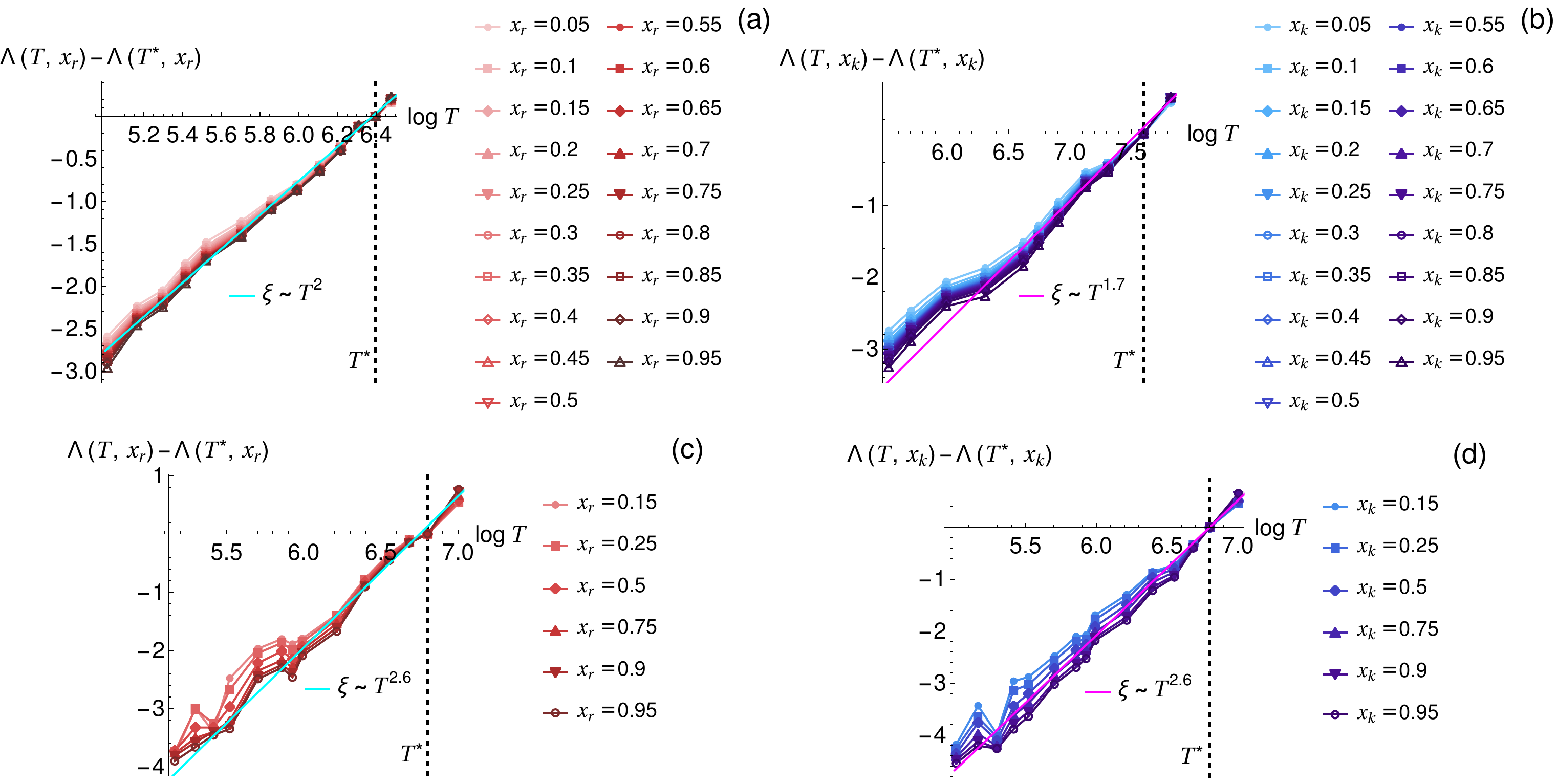}
\par\end{centering}
\caption{Plots of $\Lambda(T,x_{p})-\Lambda(T^{*},x_{p})$, where $\Lambda(T,x_{p})=\log[\xi(T,x_{p})/\xi_{0}(T,x_{p})]$,
with $p=r,k$, and $\log T^{*}$ is indicated in the figures by the
vertical dashed line. Note that the subtraction of $\Lambda(T^{*},x_{p})$
is just a trivial $x_{p}$-dependent vertical shift in $\log[\xi(T,x_{p})]$
that we applied to better see the identical slopes. (a) $V=3,\epsilon=0.45$,
for collapses in Fig.~\ref{fig:2}(f). (b) $V=1.5,\epsilon=0.375$,
for collapses in Fig.~\ref{fig:V1.5.eps-0.375}(f). (c,d) $V=2,\epsilon=0.2$,
for collapses shown in Fig.~\ref{fig:3}(d). In all figures we
show in cyan or magenta the power-laws obtained in the main text and
Fig.~\ref{fig:V1.5.eps-0.375}, for comparison. \label{fig:xi(x)}}
\end{figure*}

\begin{figure}[H]
\begin{centering}
\includegraphics[width=1\columnwidth]{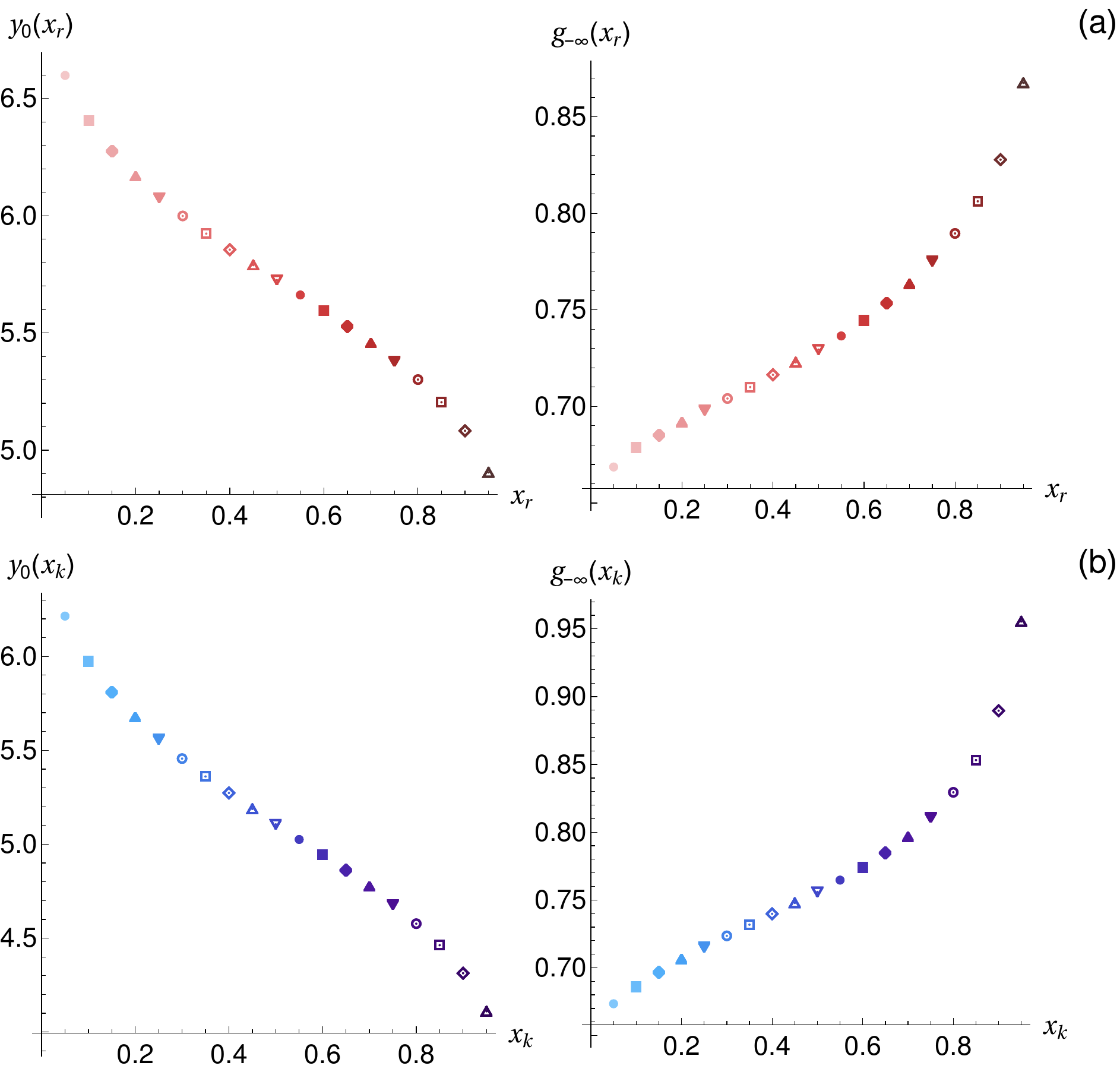}\caption{Parameters extracted from the fits to the expression $g^{(x_{p})}(y)=g_{-\infty}(x_{p})+\log(1+e^{y-y_{0}(x_{p})})$,
with $p=r,k$, for the scaling collapses in (a) Fig.~\ref{fig:2}(f);
(b) Fig.~\ref{fig:V1.5.eps-0.375}(f).\label{fig:y0_g-inf}}
\par\end{centering}
\end{figure}

\section{Details on scaling collapses}

\label{sec:scaling_collapses}

In this section, we provide details on the scaling collapses carried
out throughout the manuscript and present additional information extracted
from these collapses. 

The scaling collapse method is illustrated in Fig.~\ref{fig:scaling_collapses}
and described in detail in the figure's caption. In Fig.~\ref{fig:xi(x)},
we show plots for $\xi(T,x_{p})$ ($p=r,k$) extracted from the $x_{p}$-dependent
scaling collapses done in Figs.~\ref{fig:2},~\ref{fig:3},~\ref{fig:V1.5.eps-0.375}.
We can see that in all cases, the results depend very weakly on $x_{p}$
for large enough $T$.

Finally, in Fig.~\ref{fig:y0_g-inf}, we plot the fitting parameters
used for the scaling collapses for all $x_{p}$ in the insets of Fig.~\ref{fig:2}(f)
and Fig.~\ref{fig:V1.5.eps-0.375}(f).

\clearpage
\section{Localization-delocalization transition in Ref.$\,$ \citep{PhysRevB.103.184309} }
\label{loc_deloc_sarkar}

\begin{figure}[H]
\centering{}\includegraphics[width=1\columnwidth]{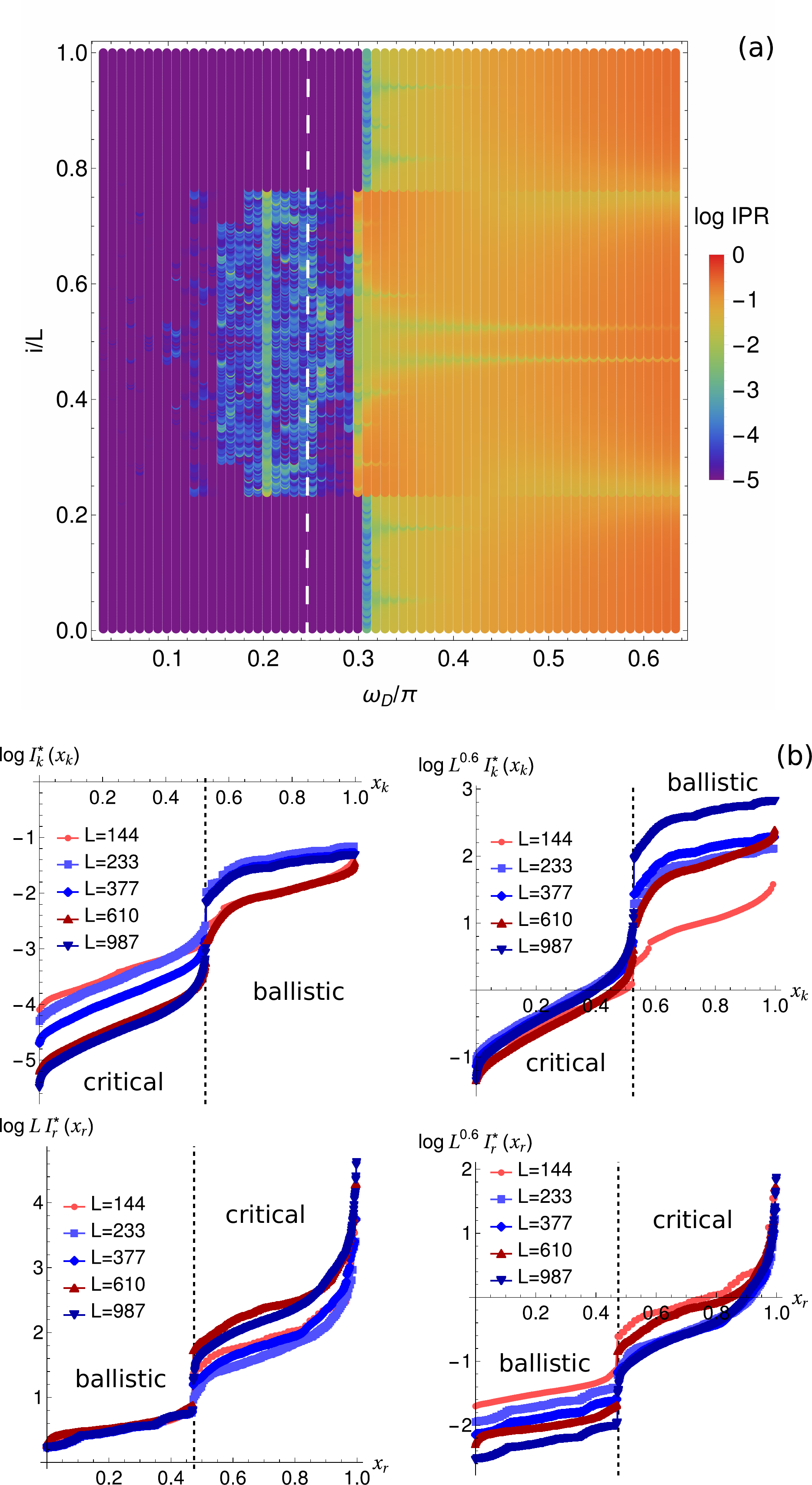}\caption{ Results for the Floquet Hamiltonian of the model in Eqs.~\eqref{eq:sarkar1},~\eqref{eq:sarkar2} introduced in Ref.$\,$\citep{PhysRevB.103.184309}, for the same parameters studied there: $V/J_{0}=0.05$, $J_0=1$ and $\tau=(\sqrt{5}-1)/2$. As in the main text, we use approximant system sizes by choosing $L=F_n$ and $ \tau=F_{n+1}/F_{n}$, where  $F_{n}$ is the n-th Fibonacci number.  (a) Color plot of the IPR for $L=144$ as a function of driving frequency $\omega_{D}$ and normalized eigenstate index $i/L$, for a random phase $\phi$ and twist $\kappa$. The eigenstates were ordered by increasing quasienergy in the interval $[-\pi, \pi[$. (b) $\mathcal{I}_k^*(x_k)$ and $\mathcal{I}_r^*(x_r)$ as defined in the main text, for $\omega_{D}=0.245\pi$ indicated in (a) (dashed white line). Up to different scaling functions for even (red points) and odd (blue points) sizes, we can clearly identify a ballistic and a critical  regime. The former and the latter correspond respectively to the extended and multifractal regimes found in Ref.$\,$\citep{PhysRevB.103.184309}.  \label{fig:S3}}
\end{figure}

In Ref.$\,$\citep{PhysRevB.103.184309}, a driven Aubry-André model was also studied, with a different driving protocol than the one studied here. The Hamiltonian in this study was given by 

\begin{equation}
\begin{array}{cc}
H(t)= & -J(t)/2 \sum_{n} c_{n}^{\dagger}c_{n+1}+\textrm{h.c.}\\
 & +\sum_{n}V \cos(2\pi\tau n+\phi)c_{n}^{\dagger}c_{n}
\end{array}
\label{eq:sarkar1}
\end{equation}

\noindent where 
\begin{equation}
J(t) =
\begin{cases}
-J_0 & t\leq T/2 \\
J_0 & t>T/2
\end{cases}
.
\label{eq:sarkar2}
\end{equation}

\noindent and $T$ is the driving period, with corresponding driving frequency $\omega_D=2\pi/T$. Since in this case there is also a square-drive protocol, the Floquet Hamiltonian can be easily obtained as in Eq.~\eqref{U_T} of the main text. In Fig~\ref{fig:S3}, we study the localization properties of the Floquet Hamiltonian of the model in Eqs.~\eqref{eq:sarkar1},~\eqref{eq:sarkar2}, for the same model parameters studied in Ref.$\,$\citep{PhysRevB.103.184309}. In Fig.~\ref{fig:S3}(a), we can clearly see that all states are localized for large enough $\omega_D$ (region of large IPR), while extended states with small IPR arise at smaller $\omega_D$, as observed in Ref.$\,$\citep{PhysRevB.103.184309}. An important question is whether this extended region is ergodic, which would go against our conjecture that ergodicity is generically not robust in the thermodynamic limit, for driven non-interacting 1D quasiperiodic systems. However, supported by the results in Fig.~\ref{fig:S3}(b), we find that the extended region is in fact a ballistic non-ergodic phase.

\end{document}